# Deep Learning Assisted Prediction of Electrochemical Lithiation State in Spinel Lithium Titanium Oxide Thin Films

*Devin Chugh[a,c], Bhagath Sreenarayanan[b], Steven Suwito[a], Ganesh Raghavendran[b], Bing Joe Hwang[a], Ying Shirley Meng[b,d], Weinien Su[a]\**

a  Graduate Institute of Applied Science and Technology, National Taiwan University of Science and Technology, Taipei, Taiwan 106
b  Aiiso Yufeng Li Family Department of Chemical and Nano Engineering, University of California San Diego, La Jolla, California 92093, United States
c  Indian Institute of Technology Jammu, Jagti, Nagrota, Jammu and Kashmir, 181221, India
d  Pritzker School of Molecular Engineering, University of Chicago, Chicago, Illinois 60637, United States

*Correspondence to wsu@mail.ntust.edu.tw

## ABSTRACT

Machine Learning (ML) and Deep Learning (DL)-based framework have evolved rapidly and generated considerable interests for predicting the properties of materials. In this work, we utilize ML-DL framework to predict the electrochemical lithiation state and associated electrical conductivity of spinel $Li_4Ti_5O_{12}$ (LTO) thin films using Raman spectroscopy data. Raman spectroscopy, with its rapid, non-destructive, and high-resolution capabilities, is leveraged to monitor dynamic electrochemical changes in LTO films. A comprehensive dataset of 3,272 Raman spectra, representing lithiation states from 0% to 100%, was collected and preprocessed using advanced techniques including cosmic ray removal, smoothing, baseline correction, normalization, and data augmentation. Classical machine learning models—Support Vector Machine (SVM), Linear Discriminant Analysis (LDA), and Random Forest (RF)—were evaluated alongside a Convolutional Neural Network (CNN). While traditional models achieved moderate to high accuracy, they struggled with generalization and noise sensitivity. In contrast, the CNN demonstrated superior performance, achieving over 99.5% accuracy and robust predictions on unseen samples. The CNN model effectively captured non-linear spectral features and showed resilience to experimental variability. This pipeline not only enables accurate lithiation state classification but also facilitates conductivity estimation, offering a scalable approach for real-time battery material characterization and potential extension to other spectroscopic datasets.

# INTRODUCTION

Raman spectroscopy has emerged as the fastest technique for determining dopant concentration in materials, delivering results in seconds compared to hours required by other methods while maintaining exceptional accuracy and spatial resolution [1]. This speed advantage, combined with non-destructive analysis capabilities, positions Raman as the premier tool for both research and industrial applications, offering measurement times that are 2-3 orders of magnitude faster than competing techniques like XPS, ARPES, and advanced electron spectroscopies [1]. The technique's rapid analysis capability stems from its direct molecular fingerprinting approach, requiring minimal sample preparation thus enabling operando monitoring of changes in dopant concentration.

Unlike electron spectroscopy methods that demand ultra-high vacuum conditions and extensive sample preparation, Raman spectroscopy operates under ambient conditions with acquisition times as short as 0.6 milliseconds for time-resolved studies [2] and typically seconds for routine characterization. This represents a transformative capability especially for battery research, where understanding dynamic lithiation processes requires both speed and precision. The speed advantage becomes stark against advanced techniques. Time-resolved Raman achieves sub-millisecond resolution—a 1000-fold improvement over conventional methods. Neale et al. [3] demonstrated sub-second to seconds acquisition for operando graphite electrode studies, enabling observation of fast electrochemical processes impossible with slower techniques. The battery characterization requires balancing speed, information depth, and accessibility. X-Ray Photoelectron Spectroscopy (XPS) provides excellent surface chemistry within 1-4 hours but analyzes only the top 5-10 nm and ultra-high vacuum requirements and complex preparation make it unsuitable for rapid screening or operando studies [4], [5] X-Ray Diffraction (XRD) offers structural information in 0.5-2 hours with operando capability but shows poor lithium sensitivity and struggles with amorphous phases and its millimeter-scale spatial resolution cannot analyze individual particles or interfaces [6], [7]. Nuclear Magnetic Resonance (NMR) excels at direct lithium detection (±1-5% accuracy) but requires 30 minutes to sometimes 12+ hours per spectrum depending on signal attentuation, making real-time monitoring impractical. Large sample quantities and specialized expertise and machinery further limit accessibility [8], [9], [10]. Neutron diffraction remains the lithium quantification gold standard (±1-3% accuracy) but requires major facility access with 3-7 day beam allocations and gram-scale samples. Total experiment time often extends to weeks, limiting use to fundamental studies rather than routine characterization [11], [12]

$Li_4Ti_5O_{12}$ undergoes a remarkable electronic phase transition from insulator to metallic conductor upon lithiation. This transition occurs through $Ti^{4+} \rightarrow Ti^{3+}$ reduction during $Li_7Ti_5O_{12}$ formation, creating percolative electronic pathways while maintaining structural stability (<0.1% lattice parameter change) [13], [14]. The complete electronic structure reorganization transforms wide bandgap semiconductor behavior (2.0-3.55 eV) to metallic conduction through partially filled Ti 3d orbitals, enabling applications in advanced batteries and neuromorphic computing where resistive switching creates voltage-driven conductive domains. Raman spectroscopy effectively monitors these electronic transitions by tracking structural vibrational mode changes that correlate with $Ti^{4+}/Ti^{3+}$ ratios.

Machine Learning (ML) has become a vital tool for spectroscopic analysis by enabling automated, precise, and efficient interpretation of spectral data. ML models leverage the relationships between spectra and sample properties to achieve tasks like classification, regression, and clustering [15]. ML algorithms find patterns in the non-linear spectroscopic data, which aids in faster classification and prediction [15]. Classical ML algorithms such as Support Vector Machine (SVM) [16], [17], [18], Linear Discriminant Analysis (LDA) [19], [20], and Random Forest (RF) [21], [22], [23] have been used for a long time for feature extraction and prediction with spectroscopic data owing to their simplicity and ease of implementation. SVM finds the optimal hyperplane that best separates the data points of different classes. Liu et al., [24] used SVM combined with Ultraviolet-Visible (UV-Vis) spectroscopy, for the discriminant analysis of the brands of the red wine. LDA is a supervised learning method that finds a linear combination of features, forming a decision boundary that separates the classes in a dataset. Silva et al., [25] used LDA combined with Infrared spectroscopy to classify the blue pen ink according to its types and brands. Random Forest builds a collection of decision trees, usually trained with the bagging (Bootstrap Aggregating) technique. Each tree is trained on a random subset of the data, and each node in a tree is split based on a random subset of features. The final prediction is made by aggregating the predictions of all individual trees through majority voting for classification. Santana et al., [26] used Random Forest combined with Infrared spectroscopy for the detection of adulteration in food samples. The SVM,

LDA, and RF are more straightforward and easier to implement; however, they are limited in finding linear relationships and are less effective for high-dimensional continuous features. These algorithms fail to generalize and accurately classify the Raman spectra especially if the dataset is smaller in size, owing to the low signal-to-noise ratio and variations in spectra due to experimental setups.

Deep learning (DL) algorithms offer superior generalization capabilities and can extract features accurately and efficiently classify them across varied experimental settings. The DL algorithms employ neural networks that capture non-linear relationships and extract high-dimensional features, which are used to make predictions [27]. Various DL architectures, including convolutional neural networks (CNNs) [28], [29], [30], residual networks (ResNets) [31], [32], recurrent neural networks (RNNs) [33], [34] autoencoders [35], [36], and generative adversarial networks (GANs) [37], [38], are used for the spectroscopic data. CNNs and RNNs are used for classification and regression tasks such as disease diagnostics (e.g., cancer detection) and material identification. GANs are particularly useful for generating synthetic data for model training, but their application with Raman Spectroscopy is minimal and still evolving. DL algorithms achieve excellent performance, but these algorithms often require a large amount of data for training, which is usually challenging to gather. The algorithms are also computationally expensive and require high-end GPUs for model training [39].

In this work, we aim to predict the physical properties of LTO thin films of different lithiation state using their Raman Spectroscopy data through various machine learning and deep learning algorithms. We collected Raman spectra from LTO thin films synthesized under varying lithiation conditions to get a diverse dataset that captures the compositional variations. The spectral data were preprocessed to remove noise and baseline fluctuations, followed by normalization to enhance feature extraction. We also employed data augmentation techniques to enhance the model's generalization and avoid overfitting. We trained the data on both machine learning and deep learning algorithms and they both showed high classification accuracy. The deep learning model (CNN) showed better generalization on unseen data and is less sensitive to noise. The models trained can efficiently and automatically estimate the lithiation percentage and other physical properties of LTO thin films using only the Raman Spectrum as an input. In this paper, the pipeline is made for analysis and prediction with LTO Raman spectral data but it can easily be modified and applied to other spectroscopic data as well.

# EXPERIMENTAL SECTION

## $Li_4Ti_5O_{12}$ (LTO) thin film synthesis

LTO thin film was deposited using Radio Frequency (RF) Sputtering with the help of a commercial $Li_4Ti_5O_{12}$ target (Toshima Manufacturing Co Ltd). The thin film was deposited on a 1 cm x 1 cm Alumina substrate (Valley Design Corp) coated on both sides with Platinum (Pt) and Chromium (Cr). The Cr interlayer was added onto the alumina substrate before sputtering Pt to improve the adhesion of Pt on the alumina Substrate. The Pt-Cr layer was deposited on the opposite side of the LTO thin film to serve as the current collector when the film is lithiated electrochemically. The sputtering process of LTO used in this work is same as from our previous work [13]. The RF power used in the process was 90 W, and the chamber pressure used was 15 mTorr to initiate plasma so that there was layer-by-layer growth of the thin film. The lost lithium in the LTO thin films during sputtering was compensated by electrochemically infusing the thin films in a coin cell using Li metal as the counter electrode. The electrolyte used for the electrochemical infusion process was LP 57, which is 1 M lithium hexafluorophosphate (LiPF6) in a 3:7 by volume mixture of ethylene carbonate (EC) and ethyl methyl carbonate (EMC). The electrochemically infused LTO film was annealed in a Box furnace at a constant temperature of 700 °C for 1 hour with a heating ramp rate of 9.5 °C/min to crystallize to the spinel structure.

## Preparation of LTO Thin Films with Different States of Lithiation

The as prepared spinel LTO films grown on Pt-Cr coated Alumina were electrochemically lithiated to various states of lithiation in the range of 0-100% in a coin cell with Li metal chip/foil as the counter electrode. The electrolyte used for the lithiation of thin films was LP 57, which is 1 M lithium hexafluorophosphate (LiPF6) in a 3:7 by volume mixture of ethylene carbonate (EC) and ethyl methyl carbonate (EMC). For 100% lithiation, the cell was lithiated to 1 V vs Li/Li+ and for all the intermediate states of lithiation, capacity cutoff was used to end the lithiation based on total capacity observed from 100% lithiation.

**Raman Spectroscopy Data acquisition**

The Raman Spectroscopy measurements were performed using Renishaw inVia Raman Microscope. The measurements were run using a 532 nm green laser source with 1800 L mm-1 grating and with 20x magnification

**Machine Learning Experimentation**

The dataset consists of 3,272 Raman spectra collected from samples with varying lithiation states ranging from 0% to 100%, divided into 11 distinct classification labels. The spectra were cropped to 150-870 cm$^{-1}$ to capture the most relevant Raman Vibrational modes of LTO. The preprocessing pipeline included cosmic ray removal through median filtering, followed by Savitzky-Golay smoothing, Asymmetric Least Squares baseline correction, and normalization between 0 and 1. Data augmentation was performed using Borderline-SMOTE to balance underrepresented classes by creating synthetic samples. The dataset was split into 70% training, 15% testing, and 15% validation sets. Further, augmentation through uniform and proportional noise addition expanded the training dataset to 17,325 spectra, which were then shuffled to eliminate bias. The analysis employed various models to process 1×350 input vectors and produce 11×1 probability vectors corresponding to lithiation states. The models included Support Vector Machine with RBF kernel, Linear Discriminant Analysis, Random Forest with 500 trees, and a Convolutional Neural Network. The CNN architecture featured three convolutional blocks with batch normalization, ReLU activation, and max-pooling, followed by dense layers. Training was conducted using Adam optimizer with a 1E-5 learning rate over 60 epochs, implementing dropout to prevent overfitting.

**RESULTS AND DISCUSSION**

Our previous work on LTO thin films showed that spinel $Li_4Ti_5O_{12}$ is an insulator (~$10^{-11}$ S/cm) but exhibits a conductivity jump of six orders of magnitude upon 4% lithiation [13]. Pristine LTO displays five Raman-active modes from Factor Group Analysis, with three major peaks: $F_{2g1}$ (~235 cm-1, O-Ti-O bending), $E_g$ (~426 cm-1, Li-O stretching in $LiO_4$), and $A_{1g}$ (~672 cm$^{-1}$, Ti-O in $TiO_6$) [40]. Two additional $F_{2g}$ peaks appear at ~270 and ~350 cm$^{-1}$ [41]. Surplus bands (~520 and ~310 cm$^{-1}$) and satellite surplus bands (shoulders at ~260, ~400, and ~750 cm$^{-1}$) arise from structural distortions and lattice defects [41]. Upon lithiation, the O-Ti-O peak red-shifts due to

$Ti^{4+}$-O to $Ti^{3+}$-O transformation, while $A_{1g}$ and $E_g$ intensities decrease and $F_{2g}$ intensity increases. At 100% lithiation, only the $F_{2g}$ peak remains; the $E_g$ peak absence indicates Li migration from tetrahedral to octahedral positions forming $Li_7Ti_5O_{12}$ [42]. The additional d-electron in $Ti^{3+}$ causes $F_{2g}$ peak broadening due to modified Ti-O force constants [42].

Many physical properties such as bandgap and electrical conductivity of the LTO thin film is highly dependent on its lithiation state. The lithiation state was predicted using machine learning models such as Support Vector Machine (SVM), Linear Discriminant Analysis (LDA), Random Forrest and Convolutional Neural Network (CNN) and then use the value to obtain the physical properties such as electrical conductivity of the LTO film. The input to the model was the Raman Spectrum collected at various points on the film and the target variable was its lithiation state. The dataset was split into 70% for training, 15% for testing, and 15% for validation sets.

The preprocessing pipeline was optimized through iterative hyperparameter tuning to obtain clean and interpretable Raman spectra for LTO thin films. Although automated spectral quality metrics exist, they are often unreliable for thin-film Raman data where peak intensities are low, noise levels vary across samples, and the important peaks such as the $F_{2g}$, $A_{1g}$ and $E_g$ must be preserved. Therefore, hyperparameters were refined primarily through visual inspection across representative spectra in the dataset. **Figure 1 (a)** shows an example of one of the representative spectra with low S/N ratio from the experimentally collected database and **Figure 1 (b)** shows its preprocessed spectrum preserving the main peaks of LTO and improving its S/N ratio. The raw spectrum exhibits high-frequency noise and baseline fluctuations that partially obscure the characteristic vibrational peaks. The processed spectrum has the peaks more clearly visible through effective preprocessing. Further, **Figure 1 (c)** shows an example of one of the representative Raman spectra with a high fluorescence background and the pre-processing strategy employed in this work can also remove this feature from the raw data and **Figure 1 (d)** shows its pre-processed Raman spectrum retaining its main peaks. The optimized preprocessing settings enhanced peak visibility while suppressing noise, enabling the model to receive consistent, high-quality inputs and thereby improving training stability and convergence, even with a limited number of spectra.

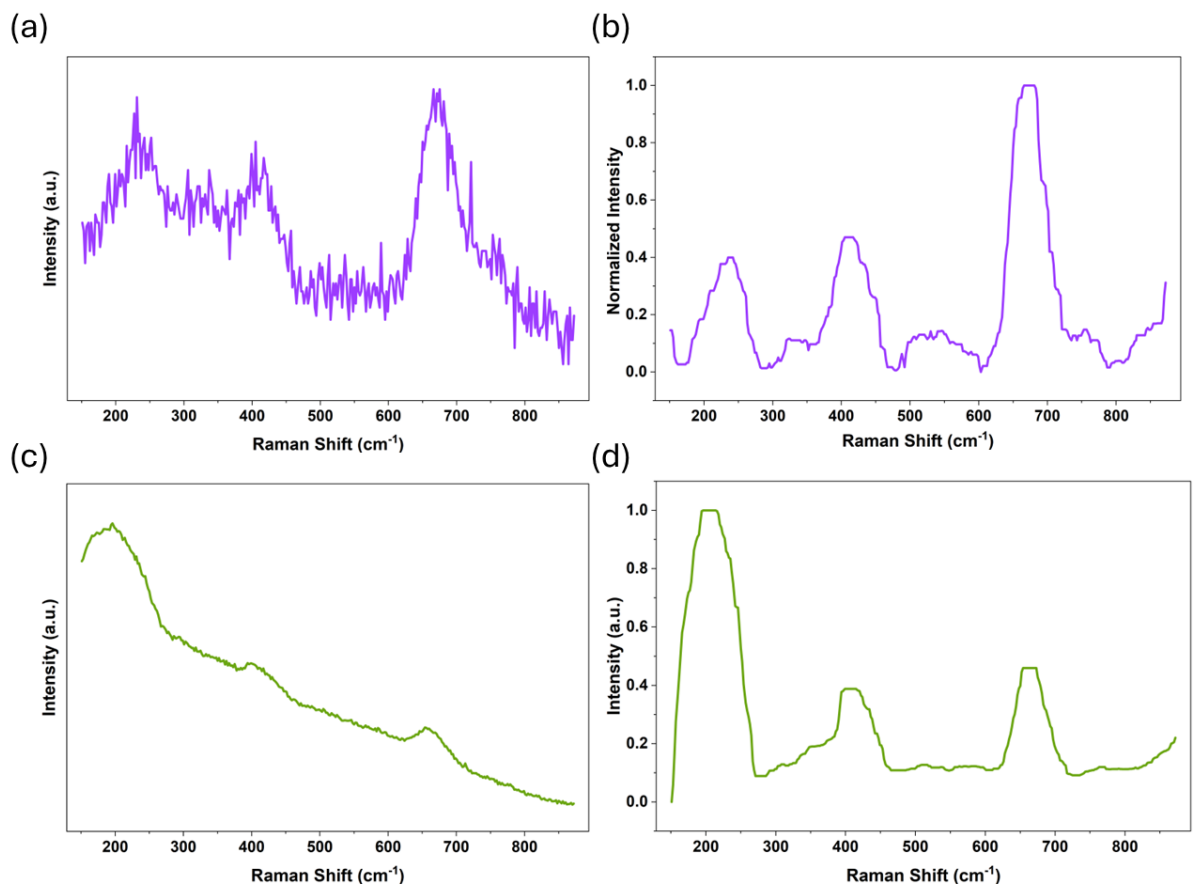

**Figure 1** (a) Representative Raw Raman Spectrum from collected database with low S/N ratio (b) its preprocessed Raman Spectrum (c) Representative Raw Raman Spectrum from collected database with possible Fluorescence background (d) its preprocessed Raman Spectrum

The data was also collected for three samples whose lithiation states were different from the samples in the dataset. These three samples were used to validate if the model can predict and accurately classify the unseen samples having the lithiation state which is not present in the training dataset. We compared these models using their confusion matrix and the accuracy of the models on training, validation, and test datasets. The confusion matrix shows how accurately a model predicts samples from each class in the test dataset. The darker color on the diagonal elements of the confusion matrix represents high accuracy of prediction for that corresponding class and vice versa. The parameters of each of the models were tuned to achieve the highest classification accuracy. We first trained the prediction models using simpler and computationally less expensive algorithms i.e. SVM, LDA, and Random Forest. The SVM model uses a Radial Basis Function (RBF) kernel, which is effective for non-linear data distributions such as spectroscopy data. The

SVM model was able to predict the samples with very distinctive features like ones with low lithiation state but showed poor accuracy on the samples having lithiation 50% or more as shown in **Figure 2**. The overall accuracy of prediction by SVM was 87%. The poor prediction by SVM was likely due to poor scalability on large dataset and high sensitivity to noisy data [43]. The LDA also showed similar results to SVM; the overall accuracy was 91%. The LDA assumes that each class follows a Gaussian distribution and that the classes are linearly separable, but the Raman data have continuous high dimension features. The LDA was also highly sensitive to noise and outliers in the data, which may be the reason for poor performance [44]. The Random Forest model was configured with 500 decision trees and trained using bagging. The Random Forest being the black-box model showed a high accuracy of over 99% on the test data set but was not able to predict well for completely unseen samples. This was primarily due to Random Forest not being able to capture the intrinsic features of the complex Raman data and may have likely overfit the data [45]. We used the trained models to make predictions for unseen samples whose lithiation states are different from those present in the training dataset. SVM, LDA, and Random Forest were unable to classify within the error range of ±10% with high confidence. While these algorithms were relatively easy to implement, they fail to capture the complex and continuous features inherent in Raman spectra. Additionally, their performance was significantly affected by the signal-to-noise ratio, which poses a challenge given the noise-prone nature of Raman spectroscopy due to experimental variability [44]. Consequently, we transitioned to a CNN-based model, as its deeper architecture is more capable of learning intricate patterns within the spectral data. The convolutional layers, which share weights and scan across the input, enable more effective feature extraction from the Raman spectra and exhibit greater robustness to noise [46].

CNN model being the most complex model required a GPU to train and took the longest time for training as compared with other models. Validation loss is used as a checkpoint and was continuously measured to see the training progression. The training and validation loss both minimized simultaneously with increasing epochs as in **Figure S2**, this show that model is able to understand the distinctive features and there is no overfitting. The overall accuracy of the CNN model is more than 99.5% as shown in confusion matrix in **Figure S1**. CNN model is also able to make prediction for the completely unseen samples with high accuracy and is the best-performing model.

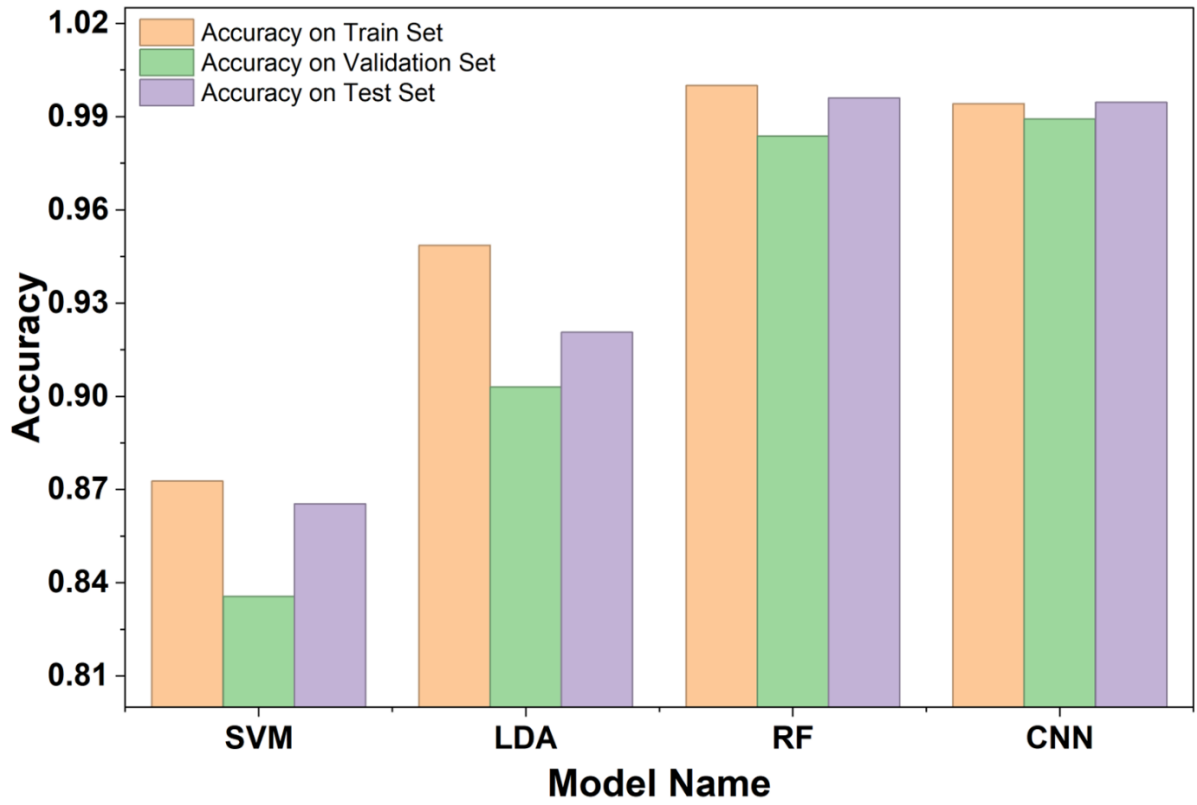

**Figure 2** Comparison of Model Accuracy on Train, Test & Validation Set

**Prediction of Lithiation State of Unseen samples**

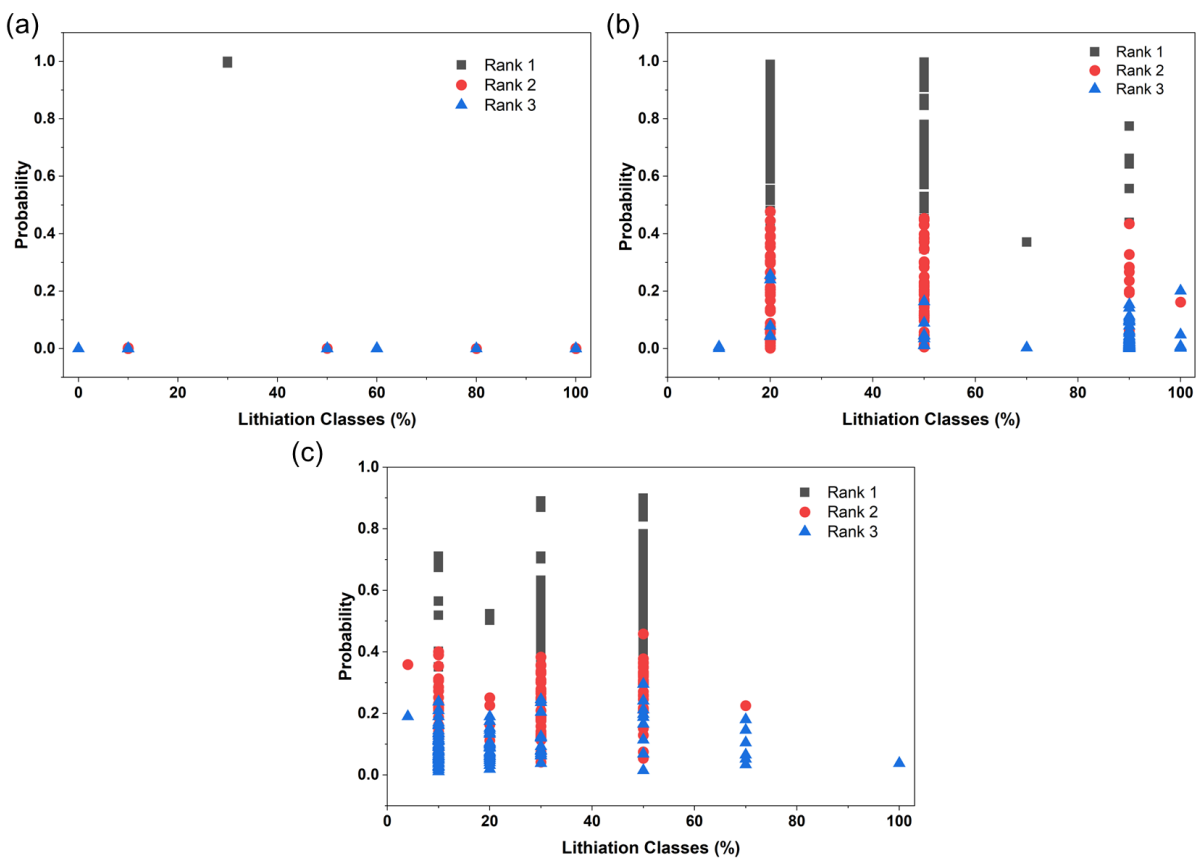

**Figure 3** Prediction results of unseen (a) Sample 1 (b) Sample 2 (c) Sample 3

The trained CNN model was used to predict the lithiation state of three unseen samples from Raman spectroscopy data, with results summarized in **Figure 3**. The visualization displays the top three ranked predictions for each sample, revealing both classification accuracy and model uncertainty through probability distributions. Sample 1 was experimentally lithiated to 25% lithiation. The model demonstrates exceptional confidence with Rank 1 predictions concentrated at 30% lithiation with probabilities approaching unity (~0.999-1.0) (**Figure 3 (a)**). The near-perfect probability scores and minimal scatter in alternative predictions indicate that the Raman spectrum contains highly distinctive features that the CNN strongly associates with the 30% class. This tight, unimodal distribution suggests the 25-30% lithiation range produces unique spectral signatures—likely specific peak positions, intensities, or widths related to the lithium insertion mechanism—and that the sample exhibits minimal heterogeneity.

Sample 2 was experimentally lithiated to 85% lithiation. The dominant predictions cluster at 90% with high probabilities (~0.95-1.0) (**Figure 3 (b)**). However, observable dispersion appears with secondary predictions at 20% and 50% lithiation. This multimodal distribution suggests several phenomena: (1) spectral ambiguity where certain features share characteristics with multiple lithiation states, (2) potential sample heterogeneity with varying local lithiation concentrations, (3) effective uncertainty quantification where the model communicates "most likely 90%, but with measurable uncertainty." The highest concentration of predictions still aligns well with the true value, demonstrating accurate primary prediction despite increased uncertainty. The Sample 3 was experimentally lithiated to 55% lithiation, exhibits the most distributed prediction pattern with significant probability mass spread across 10%, 20%, 30%, 50%, 60%, and 70% lithiation classes. Probabilities range from 0.1 to 0.9, with primary concentration around 50%. (**Figure 3 (c)**) This wide distribution reveals several insights: (1) mid-range lithiation states present greater classification challenges, likely due to less distinctive spectral features compared to composition extremes, (2) greater spectroscopic variability suggests measurement noise and (3) probability-weighted averaging across predictions would yield ~54-55%, demonstrating that discrete classification can effectively represent continuous quantities when there's minimal data available.

**CONCLUSION**

This study highlights the power of combining Raman spectroscopy with machine learning and deep learning to predict the electrochemical lithiation state and electrical conductivity of spinel $Li_4Ti_5O_{12}$ (LTO) thin films. While classical models like SVM, LDA, and Random Forest achieved moderate to high accuracy, they struggled with generalization and noise sensitivity. In contrast, the Convolutional Neural Network (CNN) model demonstrated superior performance, achieving over 99.5% accuracy and reliably predicting unseen samples. Its ability to extract complex spectral features and resist experimental variability makes it ideal for real-time material characterization. This approach offers a scalable framework for conductivity estimation and can be extended to other spectroscopic datasets and materials systems.

# AUTHOR INFORMATION

**Corresponding Author**

*(WS) E-Mail: wsu@mail.ntust.edu.tw

**Notes**

The authors declare no conflicts of interest.

# ACKNOWLEDGEMENTS

The authors gratefully acknowledge funding supported by the US Airforce Office of Scientific Research (AFOSR) (Grant FA2386-21-1-4055). The Raman spectroscopy experiments were carried out at National Taiwan University of Science and Technology (NTUST, Taipei). The electrochemical testing was performed on a battery test station donated by NeWare at University of California, San Diego (UCSD). The ML/DL models were trained using the services of Google Colab.

# AUTHOR CONTRIBUTIONS

B.S., D.C., and W.S., conceived the ideas. B.S. fabricated the samples and conducted electrochemical measurements. S.S. performed the Raman spectroscopy measurement of the samples. D.C. preprocessed the raw data and trained ML/DL models. G.R., was involved in the technical discussion of ML/DL model training and implementation. W.S., B.J.H., Y.S.M were involved in review, editing, funding acquisition, resource allocation and supervision. All authors discussed the results and commented on the manuscript.

# Supporting Information

**Deep Learning Assisted Prediction of Electrochemical Lithiation State in Spinel Lithium Titanium Oxide Thin Films**

*Devin Chugh[a,c], Bhagath Sreenarayanan[b], Steven Suwito[a], Ganesh Raghavendran[b], Bing Joe Hwang[a], Ying Shirley Meng[b,d], Weinien Su[a]\**

a   Graduate Institute of Applied Science and Technology, National Taiwan University of Science and Technology, Taipei, Taiwan 106

b Aiiso Yufeng Li Family Department of Chemical and Nano Engineering, University of California San Diego, La Jolla, California 92093, United States

c   IIT Jammu, Jagti, Nagrota, Jammu and Kashmir, 181221, India

d  Pritzker School of Molecular Engineering, University of Chicago, Chicago, Illinois 60637, United States

*Correspondence to wsu@mail.ntust.edu.tw

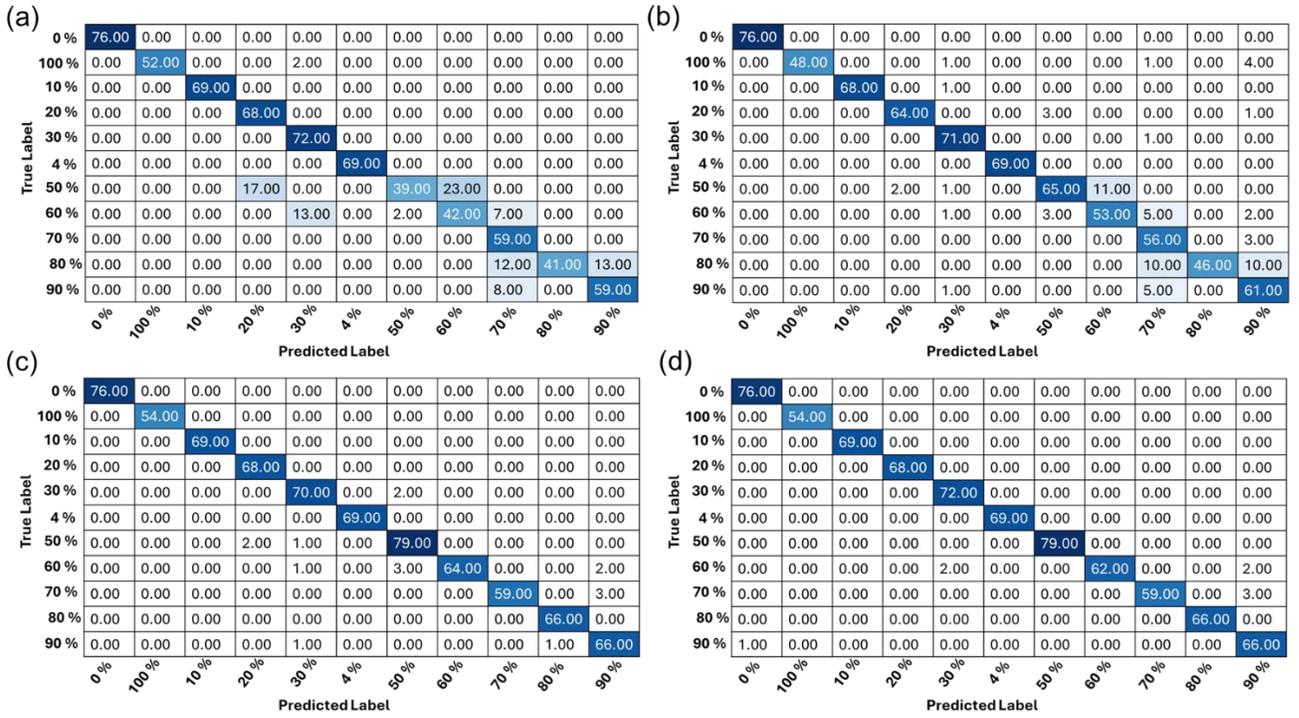

**Figure S1** Confusion Matrix of (a) SVM (b) LDA (c) RF (d) CNN

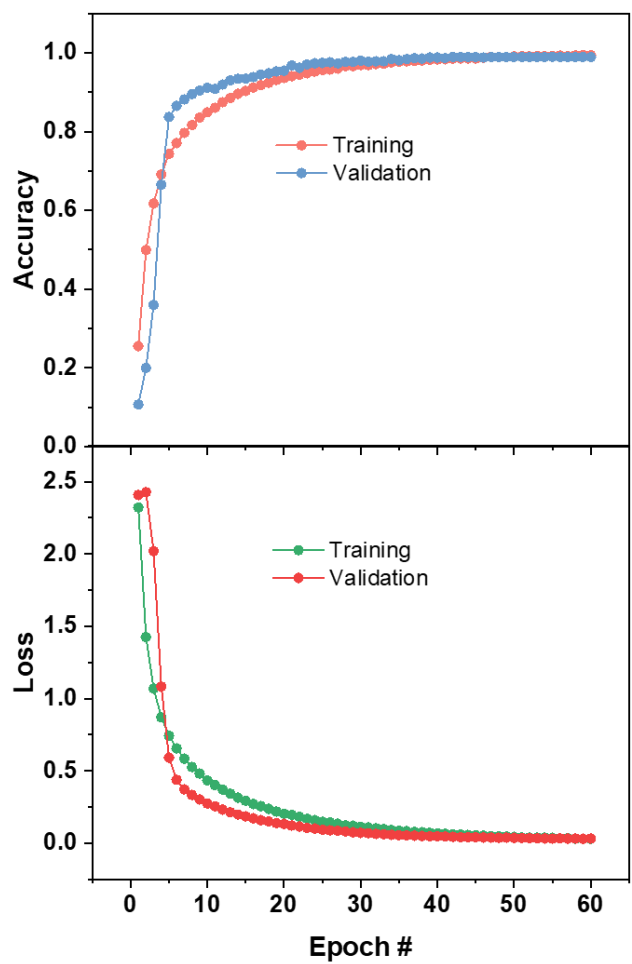

**Figure S2** Loss and Accuracy vs Epoch for CNN Model